\documentclass[10pt,conference,compsocconf]{IEEEtran}
\usepackage{times}

\usepackage{caption}
\usepackage[babel=true,english=american]{csquotes}
\usepackage[english]{babel}
\usepackage{dirtree}

\usepackage[hyphens]{url}

\ifCLASSINFOpdf
  \usepackage[pdftex]{graphicx}
\else
\fi

\usepackage{flushend}
\usepackage{orcidlink}
\usepackage{url}

\usepackage{breakurl}
\usepackage{hyperref}
\hypersetup{hidelinks,unicode,breaklinks}

\usepackage[style=ieee, backend=biber, url=false, hyperref, maxnames=5, minnames=1, maxcitenames=2, mincitenames=1, defernumbers=true]{biblatex}
\usepackage[%
babel=true, 
expansion=alltext,
protrusion=alltext-nott, 
nopatch=eqnum, 
final 
]{microtype}

\usepackage{ifpdf}
\ifpdf
\pdfoutput=1 
\pdftrue
\fi

\makeatletter
\def\ps@IEEEtitlepagestyle{
	\def\@oddfoot{\mycopyrightnotice}
	\def\@evenfoot{}
}
\def\mycopyrightnotice{
	{\footnotesize
		\begin{minipage}{0.8\textwidth}
			\centering
			Please cite as: \fullcite{selfref}.
		\end{minipage}
	}
}
\makeatother
\makeatletter
\let\blx@rerun@biber\relax
\makeatother
\usepackage[shortcuts]{extdash} 

\DeclareBibliographyCategory{selfref}
\addtocategory{selfref}{selfref}

\begin{document}

\title{\textbf{\Large Securing 3\textsuperscript{rd} Party App Integration in Docker-based Cloud Software Ecosystems}}

\author{
	\IEEEauthorblockN{~\\[-0.4ex]\large Christian Binkowski\\[0.3ex]\normalsize}
	\IEEEauthorblockA{Ostbayerische Technische Hochschule\\ Amberg-Weiden \\ Amberg, Germany\\ 	Email: {\tt c.binkowski@oth-aw.de}}
	\and
	\IEEEauthorblockN{~\\[-0.4ex]\large Stefan Appel \\[0.3ex]\normalsize}
	\IEEEauthorblockA{Siemens AG\\ \\
	Erlangen, Germany\\ Email: {\tt stefan.appel@siemens.com}}
	\and
	\IEEEauthorblockN{~\\[-0.4ex]\large Andreas A{\ss}muth\,\orcidlink{0009-0002-2081-2455}\\[0.3ex]\normalsize}
	\IEEEauthorblockA{Ostbayerische Technische Hochschule \\ Amberg-Weiden\\ Amberg, Germany \\ Email: {\tt a.assmuth@oth-aw.de}}
}

\maketitle

\begin{abstract}
Open software ecosystems are beneficial for customers; they benefit from 3\textsuperscript{rd} party services and applications, e.g. analysis of data using apps, developed and deployed by other companies or open-source communities. One significant advantage of this approach is that other customers may benefit from these newly developed applications as well. Especially software ecosystems utilizing container technologies are prone to certain risks. Docker, in particular, is more vulnerable to attacks than hypervisor based virtualisation as it directly operates on the host system. Docker is a popular representative of containerisation technology which offers a lightweight architecture in order to facilitate the set-up and creation of such software ecosystems. Popular Infrastructure as a Service cloud service providers, like Amazon Web Services or Microsoft Azure, jump on the containerisation bandwagon and provide interfaces for provisioning and managing containers. Companies can benefit from that change of technology and create software ecosystems more efficiently. In this paper, we present a new concept for significant security improvements for cloud-based software ecosystems using Docker for 3\textsuperscript{rd} party app integration. Based on the security features of Docker we describe a secure integration of applications in the cloud environment securely. Our approach considers the whole software lifecycle and includes sandbox testing of potentially dangerous 3\textsuperscript{rd} party apps before these became available to the customers.

\end{abstract}


\IEEEpeerreviewmaketitle

\section{Introduction}
Cloud computing developed within the last 10 years as the Internet is well spread all over the globe with an acceptable bandwith. With the development of Web 2.0 in the beginning of the 21st century cloud computing started spreading. Big companies like Amazon, Google or Microsoft started hosting services for companies and their applications. Cloud computing was a breakthrough technology for smaller companies as it reduced the costs of datacenter maintenance. Another important benefit is the elasticity which makes it easy for users to upscale the resources and increase the performance. Nowadays, companies  are able to push their software ecosystems easily to foreign servers instead of deploying it on their own. Another fundamental technology that was beneficial for cloud computing is the use of virtualisation. Creating multiple host systems based on shared enhanced hardware is the basis of modern cloud computing. Today it is common to run services in virtual machines on servers but in the last few years containerisation is in the focus as a new virtualisation technology. A server runs a base Linux and the services run in lightweight containers with a small OS which reduces storage costs immensely. Every container has a small host OS, the base image, which may be shared with other containers. If one user installs an Apache Server on top of the base image, Docker will add one layer to the image. Another user can install a Python environment and run scripts. This will also add another layer to the base image. This makes it easier to distribute updates to containers, as only the layer needs to be shipped to the other containers. \par
But containerisation does not only have benefits. A major problem of Docker or other containerisation services is security. The user has to adjust settings and install optional packages to create a safe environment for the use of containers.\par
One of the main advantages of Docker is Docker Hub with an immese amount of preconfigured images. However, in 2015 a study from BanyanOps showed that a lot of images uploaded to the platform are vulnerable and contain security breaches~\cite{Banyan:2015:Hub}. They tested official releases like Debian images and general images from private distributors. One third of the official images included critical vulnerabilities like the Heartbleed bug in OpenSSL, not being patched for some time. The rest of the tested images contained high and medium rated vulnerabilities. They also tested about 1700 general images supplied by 3\textsuperscript{rd} parties and the number of vulnerabilities found was even higher. The results may be interpreted as follows: even official images have weak spots and the user has to be careful when bringing these images in to his environment. \par
As the trend of cloud computing showed, many companies are creating their own cloud platforms offering different services. Some of them allow 3\textsuperscript{rd} parties to integrate applications into their ecosystems. In the context of Docker it means that a customer can push his container into the environment and interact with the provided platform services. Therefore, a security and test concept for the integration is indispensable as customers upload and process their data to the platforms. A data breach may not only result in financial consequences due to the new General Data Protection Regulation (GDPR) coming on 25th of May 2018 which forces companies to be more transparent about attacks. Any security breach then needs to be reported within 72 hours after discovery, otherwise the company has to pay up to 2\% of the worldwide turnover as a fine for a first offense~\cite{GDPR:2017:GDPR}. Beside the financial consequences the loss of trust of customers can cause an image damage of the company, too. \par
In Section~II we first present common attacks in the Docker environment, and in Section~III we discuss the Docker security features and how to increase Docker security. Section~IV describes a security concept how new containers should be able to interact with other containers. Section~V discusses a test concept for containers, and in Section~VI, we discuss our conclusions and future work.

\section{Attacks}

Researchers are finding exploits day by day and often Docker or other container environments are affected. Many different and sometimes also unconventional ways involving, e.g., social engineering, lead to damage on a Docker system. In this paper, we present mechanisms to prevent attacks using three categories of attack vectors:
\begin{itemize}
	\item Overloading the network: Denial of Service(DoS) attacks 
	\item Elevation of privileges: container breakout \& exploits
	\item Compromising the network: ARP spoofing
\end{itemize}

\subsection{Denial of Service Attacks}
Denial of Service attacks cause different effects. One purpose is bringing the host down or stop the system from operating. Denial of Service attacks are against any kind of services, interfaces or devices like the memory or cpu of the host system. Docker, not configured in the right way, is prone for DoS attacks.  \par 
As Docker refers to user namespaces every container image based on a Linux system has user ID’s (UID). For example, a UID 0 in a Docker container can relate to a UID 500 on a host system. Docker implements a constant span between the virtual UID in the container and the real UID on the host system. If a system launches 20 containers which all contain a UID 0 then every container UID will refer to the real UID 500 on the host system. This being implemented may cause Denial of Service attacks by hitting specific user limits. Different examples exist for this phenomenon described in the following paragraph~\cite{NCC:2016:CON}. \par
Every user in a Linux host system is provided a specific number of signals in order to let processes communicate with each other. A signal can stop, kill a process or transport a simple message like a number. To perform a Denial of Service attack one container tries to queue the user maximum amount for signals. As other containers share the same UID and so also refer to the same user limits they are not able to send signals anymore due to the boundaries. This may not cause a complete host takedown but some containers may freeze and will not be able to operate properly anymore. To prevent this from happening it is useful to create different users in the containers whenever possible~\cite{NCC:2016:CON}. \par
Another way to burst user limits is by increasing the number of user processes. This can bring down the Docker environment. It can be realised by creating a “fork” bomb. The father process forks itself many times and creates a lot of processes in a short time. This leads to an exceedance of the process user limit. In tests this behaviour leads to bringing down the whole Docker environment but not the host. \par
In conclusion, different ways to burst user specific limits exist. Further ways, like allocating disk space or increasing cpu usage, can cause the same effects - the host is taken down or damaged afterwards. Not only the Docker environment also the host can be attacked in this way. Attacks based on increasing disk space or cpu usage can be prevented easily by configuring the Docker environment correctly. Docker gives the user the possibility to limit the cpu usage or memory that can be used by a single container.

\subsection {Container breakout \& exploits}
Linux Kernel exploits enable attackers to break out of the container and infect the system by installing a stable backdoor and channeling malicious code. After breaking in the attacker is then able to take over control the host system or the hosted container in the environment. \par
A famous Linux kernel exploit is called "Dirty COW" which stands for "Dirty Copy On Write". In this exploit the standard user tries to write to a file that only a user with root permissions can write to~\cite{DCOW:2016:DCOW}. According to Current Vulnerabilities and Exploits (CVE)~\cite{CVED:2017:LIN}, almost half of the Linux exploits found are DoS exploits in order to bring the host system down followed by privilege elevation and information leakage. As Linux is the host system it is important to have a look at the published kernel vulnerabilities. It is also important to watch out for Docker vulnerabilites or even vulnerabilities inside the containers. This raises the potential of possible weak spots immensely. The deduction of this attack scenario is to keep the kernel, Docker, as it is a fast living environment, and software inside of containers updated and patching the system by creating new container images with updated software. 

\subsection{Address Resolution Protocol (ARP) spoofing in the container network}
ARP spoofing is popular when it comes to network sniffing and so called Man in the Middle attacks. Although Docker containers communicate over a private network they are prone to these attacks. One container could contain malicious code to spoof the private network. The following section will explain ARP spoofing shortly~\cite{ARP:2010:ARP}.\\
\begin{figure}[!h]
	\centering
	\includegraphics[width=0.45\textwidth]{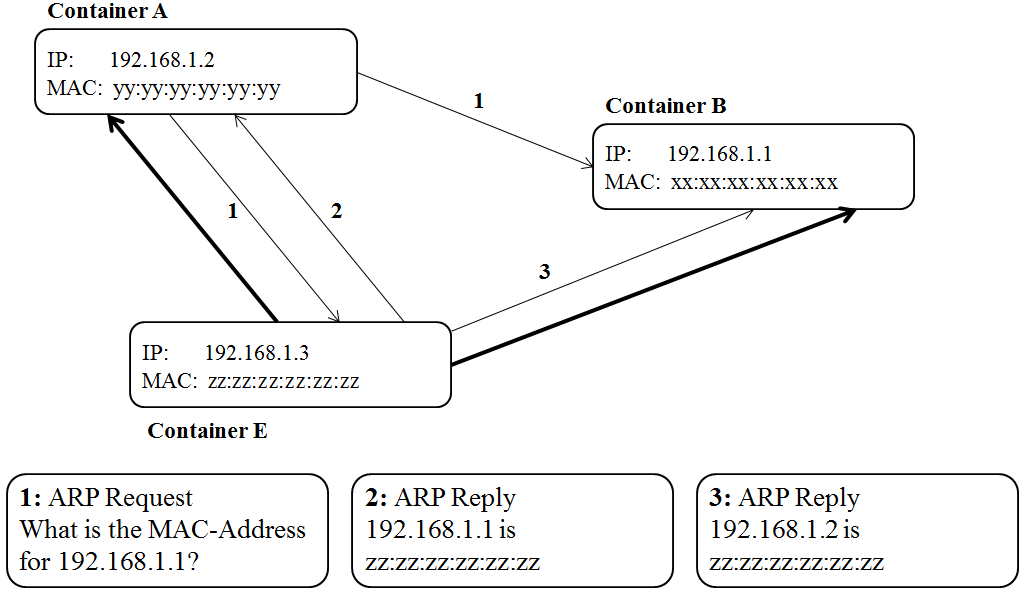}
	\caption{ARP Spoofing}
	\label{fig:arp-spoofing}
\end{figure}
Container A wants to communicate with container B and sends an ARP message which asks for the MAC address related to the IP address container A wants to communicate with. The attacking container E sends a manipulated ARP message pretending to be container B and transmits his own MAC address. Instead of establishing a connection with container  B, A is now connected with container E. All packages container A tries to send to container B now arrive at container E which can read the packages, forward them to the container B or just drop them. After spying, e.g. passwords or other credentials, container E can drop packages which leads to a loss of a connection. To perform a complete Man in the Middle attack, the malicious container has to pretend to be A when B tries to connect to container A. As both received the same MAC address they will send the packages to container E. Now he can eavesdrop on the entire communication between both containers and what attacker E just has to do is forwarding the packages between A and B. Known from public networks, e.g. cafes, spoofing and sniffing in the end can cause serious problems. \par
Using TLS as communication standard  makes it more difficult for attackers to intercept and read communications. Containers identify themselves with certificates issued by a root certificate authority. In order to prevent Man in the Middle attacks with fake or self-signed certificates, certificate pinning should be used~\cite{MAN:2009:MAN,CERTP:2015:CERTP}.

\section{Docker Security Features}
\begin{figure}[!h]
	\centering
	\includegraphics[width=0.45\textwidth]{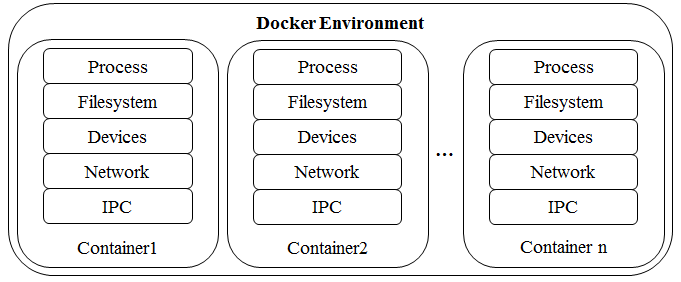}
	\caption{Namespacing in Docker}
	\label{fig:namespacing-docker}
\end{figure}

Compared to hypervisor based virtualisation a container based virtualisation tends to be more vulnerable to attacks. As a hypervisor based virtualisation has an own host OS installed, an additional layer of security is brought between the virtualised hardware and the host OS. To understand security in the Docker framework it is necessary to explain how the resources are virtualised in the Docker environment. Docker comes with some Linux specific security features that ensure the security of each container. The following two mechanisms are essential in the Docker virtualisation:

\begin{itemize}
	\item Cgroups: provides the possibility to limit the resources every container is able to access~\cite{CGROUPS:2017:CGROUPS}
	\item Namespaces: namespaces lead to a separation of spaces. In conclusion every container thinks of itself as the only container running on the system (Figure~\ref{fig:namespacing-docker})~\cite{NMSP:2014:NMSP}
\end{itemize}

Docker uses a process isolation to prevent a container from accessing the process management of other containers. The isolation is guaranteed by providing each container a unique namespace for every container, limiting the permissions and the visibility of underlying processes in other containers or the host systems. The process ID namespace isolates the process number space from the host. The process hierarchy is also a benefit for the containers as it only sees the processes in its own container or child processes. \par
The host OS and the containers must be protected from unauthorized access and modification of the file system. Thus, every container has its own filesystem and therefore can operate in its own home directory. However, some of the kernel files are not virtualised. This means, that every container shares these files with the other containers. As a Docker container is able to see the files, the system is prone to attacks like the already discussed Dirty COW attack. At least, a container is generally not able to write kernel files.\par
Another key feature of container based virtualisation is device isolation. Access to important device nodes like the physical memory or storage can cause serious damage to the host system. To prevent this from happening Docker uses the device whitelist controller which limits the access to devices for Docker. Processes are also prevented from creating new device nodes in containers.\par
Shared memories, pipes or semaphores are ways to interact with other processes. Here, Docker creates an own namespace to guarantee that every container only uses its own resources and does not communicate with processes or overwrites data in foreign shared memories. Thus, containers can't interact with processes from other containers. \par
Containers can communicate with each other only over a network connection. Every container has its own network interface including IP address, routing table, network device and stack. Docker establishes a Virtual Ethernet Bridge to communicate between container and the host system. This link can be found on the host system and is named Docker0. All hosted containers are connected to the bridge and to the eth0 interface of the container.
\begin{figure}[!h]
	\centering
	\includegraphics[width=0.45\textwidth]{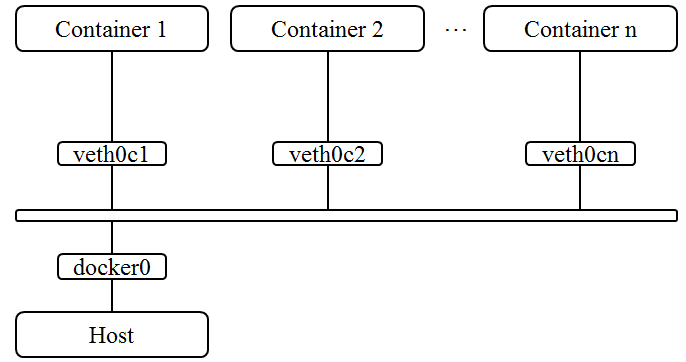}
	\caption{Docker Virtual Network}
	\label{fig:docker-vn}
\end{figure}
As Figure~\ref{fig:docker-vn} depicts, there is a stable connection between the containers and the host. With the default settings of this network setup the Docker environment is vulnerable to ARP spoofing and Man in the Middle attacks as described before. \par
To prevent DoS attacks Docker uses the cgroups functionality that allows users to configure how many resources, like cpu usage or memory, each container can access; a container is not able to consume the entire host system anymore. With the start of a container the operator can tell Docker that the container is only allowed to access a certain fraction of the available memory. Once Docker recognises a memory limit violation it will enforce the set rule.  \par
Besides of namespacing and cgroups Docker also offers other security features. It allows users to run processes in two modes:

\begin{itemize}
	\item privileged mode: using superuser permissions and no permission checks
	\item unprivileged mode: full permission checks
\end{itemize}

The following measures are introduced in order to harden the host system. All newer Linux kernels provide the possibility to assign capabilities to superusers~\cite{CAP:2017:CAP}. Capabilities are rights for super users, e.g., CAP\_NET\_ADMIN, if assigned, can be used to administrate the IP firewall or to modify the routing table. By default, many of these capabilities are disabled even if the container is running in privileged mode. In conclusion, disabling capabilities makes containers safer and the host system less vulnerable. \par
Third party applications can enhance Linux security. SELinux was created by the NSA in corporation with the Linux distributor Red Hat~\cite{SELIN:2013:SELIN}. The user or process has access to the files based on a Mandatory Access Control (MAC) system which implements rules for the user to access files. These rules can be related to categorisation or labels. Standard Linux applies a Discretionary Access Control (DAC) where decisions depend on the identity of the user.\par
SELinux offers the user a type enforcement mode that lets the user define the type of a file. The user can grant a process access to a specific type of file and doesn't have to specify the user rights for every file. Another key security feature is called Multi-category security (MCS)~\cite{SELIN:2013:SELIN}. This prohibits access to data of a foreign container. When a container is launched, the Docker daemon randomly picks a label and attaches it to every file and process that is launched or created in the container. Only containers with the consistent label can access processes or data inside the container. \par
AppArmor is a different approach to harden containers and preventing them to cause damage on the host system~\cite{APPAR:2017:APPAR}. The concept of the Linux extension refers to loading profiles in each application. Administrators can configure two modes. In enforcement mode the policies in the profile are enforced strictly. In learning mode violations are permitted but also logged on the system. The generated log file can be analysed to develop new profiles. Docker has a possibility to load profiles into containers in enforcement mode. If the user has not created any profile, it loads the default profile with less capabilities and no access to important filesystems.

\section{Security Concept}
In the following section the paper will propose a security concept which ensures a secure integration of  3\textsuperscript{rd} party applications in an existing Docker container environment. This chapter will introduce a security concept for 3\textsuperscript{rd} party app integration,  methods regarding to communication, authentication and how network analysis can improve security. \par
We want to present a concept for intra container communication, as this is crucial 3\textsuperscript{rd} party application integration. Open software platforms are an important foundation for innovative business models. However, this openess comes with risks for platform operators and customers. Especially security is an extremely important aspect; it becomes ever more challenging, when apps are not only delivered to customers, e.g., Apple/Android. Platforms allowing the execution of web-based 3\textsuperscript{rd} party apps need to be secured against potentially malicious components. On the other side, it is very important to open the platform for other developers as it will only grow with external input. \par
Cloud was an enabler for ecosystems in recent years. Developers easily pushed their deployments on servers in cloud farms and could scale the applications without maintaining a big datacenter on their own. That is not only a benefit for the developer, also the customer profits by a growing variety of Software as a Service (SaaS) offers. \par
A number of researchers already have demonstrated that Docker has some serious security concerns. They focused their research on container hardening to minimize the number of weak spots of a Docker container. Others refer to hardening the host and its services in order to reduce the attack surface for Docker containers~\cite{BenchCIS:2017:BenchCIS}. \par
We offer a solution for the third party app integration in the docker environment and are adding another layer of security in addition to hardening the host and containers. Our solution focuses on the services architecture~\cite{MICRO:2016:MICRO}. The ecosystem is based on small services that interact with a Representional State Transfer (REST) Application Programming Interface (API)  concept. An API Gateway routes the requests to the demanded services. As the services are not bound to programmining languages it is easy for 3\textsuperscript{rd} parties to develop a service in a Docker container and integrate it into the ecosystem. If one container is down, the platform still can be up and running because the services are not centralised on a single host. 

\subsection{Request Forwarding}
As described in chapter 2 all containers share the same network interface which allows every container to communicate within the network. Once a new container is integrated into the platform, it could interact with the other containers without restrictions. To prevent this, a proxy container is introduced into the network. \par
This prohibits two containers from exchanging messages directly. The functionality of a network proxy is described briefly:

\begin{itemize}
	\item The client requests a service.
	\item The proxy receives the request and forwards the message to the original server.
	\item The response of the server will be processed by the proxy which will forward the response to the client in the end.
\end{itemize}

A proxy depicts a Man in the Middle which can control the network traffic. 3\textsuperscript{rd} party application in the environment should only be allowed to access a selected services that are available in the environment. In this concept we would like to introduce two proxies for different use cases. 
\begin{figure}[!h]
	\centering
	\includegraphics[width=0.45\textwidth]{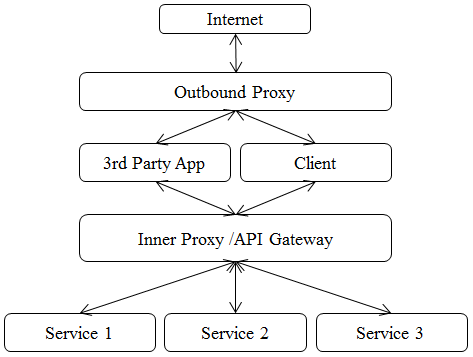}
	\caption{Proxy Setup}
	\label{fig:proxy-setup}
\end{figure}
There is one proxy, also called API Gateway, that will handle the requests regarding to the backend services of the platform. The second proxy is used for the outbound traffic towards the Internet. Both proxies can implement rules to refuse or drop connections which a container wants to establish. Although this makes the communication more complex it adds a security layer between the backend services and the 3\textsuperscript{rd} party and also a additional layer between the Internet and the foreign application. The communication and benefits of this setup is described in the sub paragraphs Logging and Whitelisting. Another benefit of a proxy is that the operator can ask for authentication and authorisation before granting access to the services. Only containers with the correct credentials or registration are able to use the proxy and thus the services in other containers This is a first step to isolate services from direct requests in order to make them more secure. \par

\subsection{Encrypted Communication}

As described in Section III, ARP spoofing or Man in the Middle attacks are a risk. As mitigation we propose to implement secure connections by using Transport Layer Security (TLS)~\cite{TLS:2008:TLS}. TLS is a protocol to establish an encrypted and authenticated connection. During the handshake, client and server perform a key exchange. We recommend to use strong ciphersuites to mitigate the risk of an attacker being able to crack the keys. Both, client and server, identify themselves via X.509 certificates issued by a root certificate authority (CA). By signing the key exchange messages with the X.509 certificate, server and client prove that they are the sender of these messages. As the platform can't be reached directly from the Internet (Figure~\ref{fig:proxy-setup}) an internal Public Key Infrastructure can be deployed. After the handshake the communication between server and client is encrypted, and message authentication codes additionally provide authenticity. \par
ARP spoofing attacks are based on rerouting packages to a wrong MAC address. If an attacker performs ARP spoofing in a network which deploys TLS, the attacker will receive encrypted messages, which he can't decrypt without the correct key. A successful spoofing attack now requires using fake certificates to pretend being a different container. Those fake certificates can only be issued, if the root CA was attacked. In conclusion by using TLS and certificates the gain of security is immense. A container can't easily spy on credentials or authentication keys, which might elevate his rights in the system to gather confidential information, e.g. customer data. Another extension to prevent Man in the Middle attacks is certificate pinning. When certificate pinning is deployed, the container requests a server connection and the server will pin the transmitted certificate to the container that sent the request. The server then only accepts connections when the submitted certificate matches the pinned certificate. An attacker could add a self signed or other certificates via a proxy and so bypass the TLS security mechanisms. Certificate pinning can prevent this scenario as the server does not trust all connections that would be verified by a root certificate. This extension was added to the handshake protocol as CA's lost trust due to attacks and fake certificates.   

\subsection{Logging}

The API Gateway protects the services from direct requests from a 3\textsuperscript{rd} party app. The Gatway reroutes the requests to the required platform service. But the proxy besides rerouting can extend the functionality and log incoming requests. Analysis of requests and the HTTP/HTTPS status codes can help to identify malicious behaviour. We want to present three different ways for a container to act in a malicious way.\par
A correct request leads to HTTP status code 200. If a container sends a request with a wrong syntax, the server will respond with status code 400. When the inquiring container uses a wrong access token or is not registered to the service, the service will respond with a status code 401. A container can also send too many requests in order to bring it down (DoS attack), the API Gateway can measure that as well. Thus the operator can create statistics on container behavior and identify suspicious activities. For example, a user tries to access the weather API and the database API with no permission. After a certain time the API gateway detects that the container sends too many requests without the correct permission. The gateway can automatically evaluate the received status codes and ban the container from the network if the numbers of unauthorized requests is too high. Besides unauthorized and wrong requests, the number of requests enable the gateway to identify DoS attacks. If a single container sends too many request in a short time period, e.g. calling the weather API for 1000 times in five seconds in order to bring the service down, the API gateway can identify the malicious container and reduce its bandwith or block requests and isolate it from the platform services~\cite{CERTP:2015:CERTP}.\par
As logging data is highly valuable for digital forensics and analysis it is important to store the collected data in a secure way. Right now, we are also working on a concept to mitigate the risk of data loss or manipulation. In general, a system based on the approach of Weir and A{\ss}muth, which includes Message Authentication Code (MAC) chains and secret sharing, may be a first step towards securing monitoring data~\cite{MONITOR:2017:MONITOR}.

\subsection{Whitelisting/Blacklisting}

Also the outgoing traffic to the Internet (Figure~\ref{fig:proxy-setup}) from containers can indicate attacks. Some commands inside the container require additional downloads from the Internet. Downloaded content from the Internet can include malicious software or exploit code that tries to cause damage in the platform. For instance, a container might download new kernel exploits to elevate permissions and damage the host system. To prevent unbridled downloads, the platform operator has to create mechanisms to keep the risk of malicious downloads small. For this reason we propose the outbound proxy in Figure~\ref{fig:proxy-setup}. One possible measure is to define guidelines for developers of 3\textsuperscript{rd} party containers on allowed traffic. In case of a detected rule violation the container can be banned from the network immediatly. In addition the administrator can create a whitelist of IP adresses in the proxy which every container is allowed to reach out to. Every IP address or host name not included in the whitelist is blocked. The platform operator can ensure that containers only download software from trusted sources like IP addresses from specific countries or sources. So locating the IP address of a request going to the Internet can help to improve security in the network.

\subsection{Application Authorisation}

Once a request has passed the API gateway it will be directed to the service. The container has to prove that it has the permission to use the service which brings in authentication. A common standard for securing API's is OAuth2~\cite{OAUTH:2012:OAUTH}. Instead of submitting username and password to the server to gain access to services, OAuth2 follows a token-based approach for this. Once a container wants to access a service he needs to request authorisation at the resource owner. This step is required to identify the container can be seen as a replacement for identifying through username and password in every request. The service operator allows access to the service and therefore will send an authorisation string to the client. Every time the container wants to access the service he has to reach out to the authorisation server with the authorisation string. The server will compute an access token and send it back to the container. In further service API requests the client has to submit the token to the server in order to receive the required information.  Access tokens have a certain time span. If the access token expires, the client has to request another token. As described before, the user doesn't need to submit username and password to the server as he only sends a randomly chosen access string. This leads to anonymity as an attacker can't refer to the specific user if the connection is eavesdropped. Combined with TLS in connection establishing the system is secured. Now it is hard for attackers to spy on credentials. If the attacker is able to manage an attack against TLS, the distributor can issue a new identification string and can make the cracked authorisation string invalid.\par
In summary, the platform services are now secured from direct access through the API gateway. In addition the logging functionality enables the operator to identify malicious activities. As the Docker network is prone to ARP spoofing and Man in the Middle attacks the implemented TLS hardens the network communication and hard for a container to spy on data, e.g credentials. With OAuth2 we anonymise the user requests and secure the API's from unauthorized access. The outbound proxy filters requests towards the Internet and therefore protects the platform from downloaded exploits. These security features help an operator to mitigate the risk of an attack.

\section{Test Concept}

To mitigate the risk of damage in the container ecosystem, it is indispensable to test the components before integrating them into the system. In the software lifecycle test is an important step before the final product release. It is common to test software to find bugs and to improve software quality but ensuring the security of products has also become more important in recent years. In this section we will propose three different methods to improve security by testing the software:

\begin{itemize}
	\item Static Code Analysis
	\item Dependency Checks
	\item Sandbox Testing
\end{itemize} 

\subsection{Static Code Analysis} 
Static code analysis can take place before the functionality of the software is tested~\cite{SCA:2017:SCA}. The purpose of the test is improving software quality by finding bad coding styles or duplicate code. But not only from the quality perspective it is important to perform a static code analysis. This process also helps to improve the security of the written software. Test tools are able to find vulnerabilities like race conditions, buffer overflows or memory leaks before a attacker can exploit them. \par
But static code analysis has to be performed in the providing company, as no company will voluntarely share their source code and know-how with the hoster of the platform. A vendor of a platform has to create guidelines which tell the developers what security standards must be implemented and what kind of tests need to be performed before application submission. One important part of testing will be described in the next section. 

\subsection{Dependency Checks}
\begin{figure}[!h]
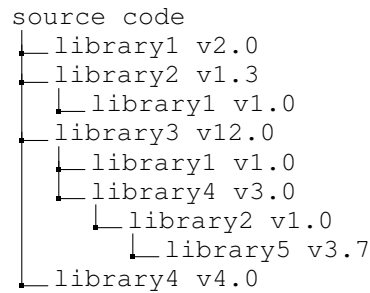

	\dirtree{%
	.1 source code.
	.2 library1 v2.0.
	.2 library2 v1.3.
	.3 library1 v1.0.
	.2 library3 v12.0.
	.3 library1 v1.0.
	.3 library4 v3.0.
	.4 library2 v1.0.
	.5 library5 v3.7.
	.2 library4 v4.0.
	}
	\caption{Example Dependency Tree}
	\label{fig:dep-tree}
\end{figure}

Most software make use of libraries. The libraries themself can also use other libraries which creates a tree of library dependencies. The scenario, shown in Figure~\ref{fig:dep-tree}, helps to illustrate the possible weak spot in the source code.

For instance, the developer always uses the updated versions while the other layers may include outdated versions of the libraries. Those old versions may be prone to vulnerabilities which an attacker can exploit. When the application then runs in a container, the container could be inherited by a 3\textsuperscript{rd} party like an attacker, but also the vendor of the application can use these unstable or vulnerable libraries to create backdoors on purpose in order to perform malicious activities. Hence, it is important to check the dependencies of the source code before the integration of the container. As stated previously it might be difficult to get the original source code, so the developer could hand over some kind of "header file", that reports which libraries are used in the application. Then the container hoster can continously check the header files against a vulnerability database and send out reports if the used libraries are exploitable. The developer could be forced to update his application within a certain timespan or it might be excluded from the platform. In order to keep the platform stable and secure, the vulnerability database needs to be updated regularly.

\subsection{Sandbox Testing}
The final step before the integration of the container into the platform is to perform a sandbox test. Sandbox testing is a good way to evaluate how the application interacts with the provided services. The hoster creates a development space next to the "Live Space", which can be used to test the applications and how they react in different scenarios. \par
Sandbox testing can also be helpful for the developer to see how the application works in a live system. The developer cannot cause any damage to a live system as he tests the software in a different space. This test is meant to reduce possible downtime of a host system.\par
Once an application doesn't show any malicious activities and worked fine in the sandbox environment, it can be integrated into the live system. However, a sandbox has to be developed, equipped with analysis tools in order to monitor behavior correctly, maintained and hosted. 
To reduce effort, a smaller demo environment, which represents only the critical interfaces of the platform may be helpful to monitor the behavior of a container.\par
The sandbox can not only be used to identify possible malicious containers, it also helps to improve the security of the system as it reveals possible vulnerabilities. Some of the containers are maybe not meant to be malicious on purpose and therefore show the behavior due to bugs in the software. So the operator has to distinguish between an attacker that wants to harm the system and a developer who did not write the software not properly. All in all, a sandbox environment can strengthen the security of a platform especially when it allows 3\textsuperscript{rd} parties to develop and deploy apps into the ecosystem.

\section{Conclusion \& Future Work}
Our approach considers the whole software lifecycle and also includes sandbox testing of potentially dangerous 3\textsuperscript{rd} party apps before these became available to the customers. We presented different techniques in order to secure the system from the described attacks. The described architecture and security features help the operator to decouple and secure services from the 3\textsuperscript{rd} party app. Hence, the foreign app is not able to access the services without permission and control. Testing can be tough as the app developer will not supply the operator with the source code. Developer guidelines can help to streamline the process and ensure high quality \par
We are currently working on a proof of concept system to get these concepts into practice. We intend to build backend services and protect them with an API Gateway as depicted in Figure~\ref{fig:proxy-setup}. Afterwards we will attack this demo system in various ways~-- our results will then be published. 
\\

\renewcommand*{\bibfont}{\footnotesize}
\setlength{\labelnumberwidth}{0.45cm}
\printbibliography[notcategory=selfref]

\end{document}